\documentclass[%
reprint,
superscriptaddress,
amsmath,amssymb,
aps,
pra,
longbibliography,
]{revtex4-1}

\usepackage{braket}

\usepackage{subfigure} 
\usepackage{graphicx}
\usepackage{dcolumn}
\usepackage{bm}

\usepackage{xcolor} 
\usepackage{hyperref}
\hypersetup{
colorlinks,
linkcolor={red!50!black},
citecolor={blue!50!black},
urlcolor={blue!80!black}
}

\usepackage{tikz}
\usetikzlibrary{patterns}
\usepackage{braket}
\usetikzlibrary{backgrounds,shadows.blur,fit,decorations.pathreplacing,shapes}

\usepackage[frozencache]{minted}  
\usepackage[skins]{tcolorbox}
\newtcolorbox{codebox}{enhanced,width=.5\textwidth,center upper,drop fuzzy shadow southeast,boxrule=0.4pt,sharp corners,colframe=black,colback=white}

\usepackage{xspace}

\newcommand{\figref}{Figure\xspace}
\newcommand{\tabref}{Table\xspace}

\newcommand{\projectq}{ProjectQ\xspace}
\newcommand{\quest}{QuEST\xspace}
\newcommand{\questGPU}{QuEST-GPU\xspace}
\newcommand{\qhipster}{qHipster\xspace}
\newcommand{\quantumpp}{Quantum++\xspace}

\newcommand{\anaconda}{Anaconda\xspace}
\newcommand{\python}{Python\xspace}
\newcommand{\clang}{C\xspace}
\newcommand{\cpplang}{C++\xspace}

\newcommand{\icc}{ICC\xspace}
\newcommand{\gcc}{GCC\xspace}
\newcommand{\openmp}{OpenMP\xspace}
\newcommand{\mpi}{MPI\xspace}
\newcommand{\cuda}{CUDA\xspace}
\newcommand{\avx}{AVX}
\newcommand{\simd}{SIMD\xspace}

\newcommand{\arcus}{ARCUS Phase-B\xspace}
\newcommand{\archer}{ARCHER\xspace}

\newcommand{\linux}{Linux\xspace}

\newcommand{\llc}{LLC\xspace}
\newcommand{\cpu}{CPU}
\newcommand{\gpgpu}{GPGPU}
\newcommand{\gpu}{GPU\xspace}
\newcommand{\numa}{NUMA\xspace}

\usepackage{soul}
\newcommand{\newedit}[1]{#1}

\begin{document}


\title{QuEST and High Performance Simulation of Quantum Computers}

\author{Tyson Jones}
\affiliation{Department of Materials, University of Oxford, Parks Road, Oxford OX1 3PH, United Kingdom}
\author{Anna Brown}
\author{Ian Bush}
\affiliation{Oxford e-Research Centre, Department of Engineering Science, University of Oxford, Keble Road, Oxford OX1 3PH, United Kingdom}
\author{Simon Benjamin}
\affiliation{Department of Materials, University of Oxford, Parks Road, Oxford OX1 3PH, United Kingdom}

\date{\today}


\begin{abstract}
We introduce \href{https://quest.qtechtheory.org/}{\quest}, the Quantum Exact Simulation Toolkit, and compare it to \projectq~\cite{projq_whitepaper}, \qhipster~\cite{qhipster} and a recent distributed implementation~\cite{distributed_quantum_plus_plus} of \quantumpp~\cite{quantum_plus_plus}.
\quest is the first open source, \openmp and \mpi hybridised, \gpu accelerated simulator {of universal quantum circuits}.
{Embodied as a C library, it is designed so that a user's code can be deployed seamlessly to any platform from a laptop to a supercomputer.}
{QuEST is} capable of simulating generic quantum circuits of general single-qubit gates and multi-qubit controlled gates, \newedit{on pure and mixed states, represented as state-vectors and density matrices, and under the presence of decoherence}.
Using the \arcus and \archer supercomputers, we benchmark \quest's simulation of random circuits of up to 38 qubits, distributed over up to 2048 compute nodes, each with up to 24 cores. We directly compare \quest's performance to \projectq's on single machines, and discuss the differences in distribution strategies of \quest, \qhipster and \quantumpp. 
\quest shows excellent scaling, both strong and weak, on multicore and distributed architectures.
\end{abstract}

\maketitle




\section{Introduction}

Classical simulation of quantum computation is vital for the study of new algorithms and architectures. As experimental researchers move closer to realising quantum computers of sufficient complexity to be useful, their work must be guided by an understanding of what tasks we can hope to perform. This in turn means we must explore an algorithm's scaling, its robustness versus errors and imperfections, and the relevance of limitations of the underlying hardware. 
{Because of these requirements simulation tools are needed on many different classical architectures; while a workstation may be sufficient for the initial stages of examining an algorithm, further study of scaling and robustness may require more powerful computational resources.}
{Flexible, multi-platform supporting simulators of quantum computers are therefore essential.}

{Further it} is important these simulations are very efficient since they are often repeated many times, for example to study the influence of many parameters, or the behaviour of circuits under noise. 
But it is expensive to exactly simulate a quantum system using a classical system, since a high-dimensional complex vector must be maintained with high fidelity. Both the memory requirements, and the time required to simulate an elementary circuit operation, grow exponentially with the number of qubits.
A quantum computer of only 50 qubits is already too large to be comprehensively simulated by our best classical computers~\cite{49_qb_sim_intel}, and is barely larger than the 49 qubit computers in development by Intel and Google~\cite{intel_49qb,google_aim_for_49qb}.
To simulate quantum computers even of the size already experimentally realised, it is necessary that a classical simulator take full advantage of the performance optimisations possible of high performance classical computing.

{
It is also equally important that the research community have access to an ecosystem of simulators. Verfication of complex simulations is a non-trivial task, one that is much eased by having the facility to compare the results 
of simulations performed by multiple packages.
}

{The number of single compute node generic}~\cite{microsoft_liquid,pyquil_whitepaper,microsoft_q_sharp} {and specialised}~\cite{graph_based_qc_simmer,qtorch_tensor_qc_simmer,clifford_t_polynomial_algorithm,quantum_network_simmer} {simulators is rapidly growing.} 
{However}
despite many reported distributed simulators ~\cite{apparently_massive_parallel_qc_sim,first_parallel_qc_simmer,parallel_qc_simmer_noisy,qhipster,haener_45qb_sim,qx_high_performance_qasm_simulator,distributed_quantum_plus_plus,big_64qb_mpi_sim_approx} and proposals for \gpu accelerated simulators~\cite{haener_45qb_sim,gpu_qc_sim_proposal,distributed_gpu_qc_sim_proposal,gpu_qc_sim_review,shor_sim_on_GPU}, \quest is the first open source simulator available to offer {\textit{both} facilities}, and the only simulator to offer {support on all hardware plaforms commonly used in the classical simulation of quantum computation}


\section{Background}


\subsection{Target Platforms and Users}
{Simulations of quantum computation are performed on a wide variety of classical computational platforms, from standard laptops to the most powerful supercomputers in the world, and on standard CPUs or on accelerators such as GPUs. Which is most suitable for the simulation of a given circuit will depend upon the algorithm being studied and the size of the quantum computer being modelled. To date this has resulted in a number of simulators which typically target one, or a small number, of these architectures. While this leads to a very efficient exploitation of a given architecture, it does mean that should a research project need to move from one architecture to another, for instance due to the need to simulate more qubits, a different simulation tool is required. This may require a complete rewrite of the simulation code, which is time consuming and makes verification across platforms difficult. In this article we describe QuEST which runs efficiently on \textit{all} architectures typically available to a researcher, thus facilitating the seamless deployment of the researcher's code. This universal support also allows the researcher to easily compare the performance of the different architectures available to them, and so pick that most suitable for their needed simulations.}

{In the rest of this section we shall examine the nature of the architectures that are available, cover briefly how codes exploit them efficiently, and show how QuEST, the universal simulator, compares with the more platform specific implementations.}


\subsection{Simulator Optimisations}

Classical simulators of quantum computation can make good use of several performance optimisations.

For instance, the data parallel task of modifying the state vector under a quantum operation can be sped up with single-instruction-multiple-data (\simd) execution. \simd instructions, like Intel's advanced vector extensions (\avx), operate on multiple operands held in vector registers to concurrently modify multiple array elements~\cite{intel_avx_whitepaper}, like state vector amplitudes.

Task parallelism can be achieved through multithreading, taking advantage of the multiple cores found in modern \cpu s. 
Multiple \cpu s can cooperate through a shared \numa memory space, which simulators can interface with through \openmp~\cite{openmp_versions}.

Simulators can defer the expensive exchange of data in a \cpu{}'s last level cache (\llc) with main memory through careful data access; a technique known as cache blocking~\cite{cache_blocking}. Quantum computing simulators can cache block by combining sequential operations on adjacent qubits before applying them, a technique referred to as \textit{gate fusion}~\cite{qhipster,haener_45qb_sim}.
For instance, gates represented as matrices can be fused by computing their tensor product.

Machines on a network can communicate and cooperate through message passing. Simulators can partition the state vector and operations upon it between distributed machines, for example through \mpi, to achieve both parallelisation and greater aggregate memory. Such networks are readily scalable, and are necessary for simulating many qubit circuits~\cite{haener_45qb_sim}.

With the advent of general-purpose graphical processing units (\gpgpu{}s), the thousands of linked cores of a \gpu can work to parallelise scientific code. Simulators can make use of NVIDIA's compute unified device architecture (\cuda) to achieve massive speedup on cheap, discrete hardware, when simulating circuits of a limited size~\cite{gpu_qc_sim_review}.
We mention too a recent proposal to utilise multi-\gpu nodes for highly parallel simulation of many qubit quantum circuits~\cite{distributed_gpu_qc_sim_proposal}.


\subsubsection{Single node}


\projectq is an open-source quantum computing framework featuring a compiler targeting quantum hardware and a \cpplang quantum computer simulator behind a \python interface~\cite{projq_whitepaper}. 
In this text, we review the performance of its simulator, which supports \avx{} instructions, employs \openmp and cache blocking for efficient parallelisation on single-node shared-memory systems, and emulation to take computational shortcuts \cite{projq_emulator_abilities}.

\quest is a new open source simulator developed in ISO standard conformant \clang \cite{ISO_C_99}, and released under the open source MIT license. Both \openmp and \mpi based parallelisation strategies are supported, and they may be used together in a so-called hybrid strategy. This provides {seamless} support for both {single-node} shared-memory and distributed systems. \quest also employs \cuda for \gpu{} acceleration, 
and offers the same interface on single-node, distributed and \gpu{} platforms.
Though \quest does not use cache blocking or emulation, we find \quest performs equally or better than \projectq on multicore systems, and can use its additional message-passing facilities for faster and bigger simulations on distributed memory architectures.

{
ProjectQ offers a high-level Python interface, but can therefore be difficult to install and run on supercomputing architectures, though containerisation may make this process easier in future}~\cite{projectqOnSupercomputersGuide,projq_cache_anomaly_haener}.
{Conversely, Quest is light-weight, stand-alone, and tailored for high-performance resources - 
its low-level C interface can be compiled directly to a native executable and run on personal laptops and supercomputers.
} 

\begin{figure}[t]
\centering
\includegraphics[width=.49\textwidth]{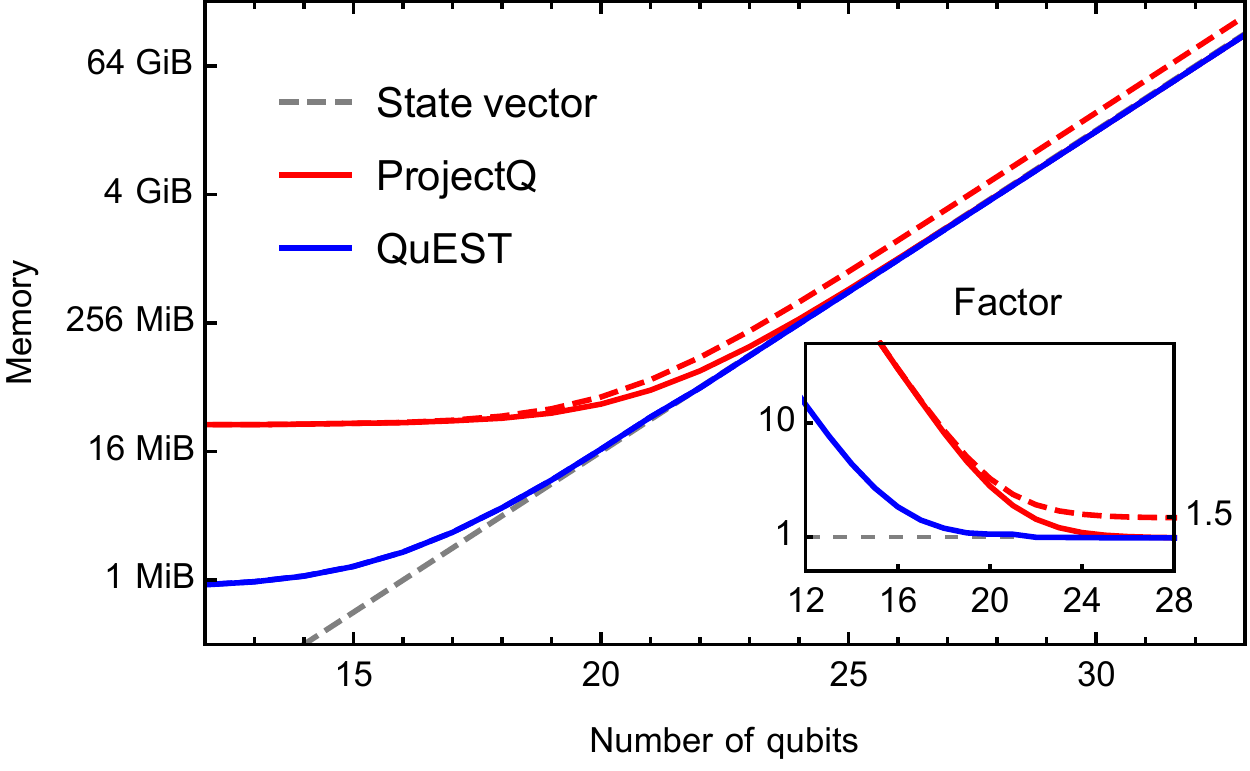}
\caption{Memory consumption of \quest's \clang and \projectq's \python processes, as reported by \linux's \texttt{/proc/self/status} during random circuit simulation on a single 256\,GiB \arcus compute node. Full and dashed lines show the typical and maximum usage respectively, while the gray dashed line marks the memory required to store only the state-vector (in double precision). The subplot shows the ratio of total memory consumed to that by only the state-vector.}
\label{fig:quest_projq_memory}
\end{figure}

Both \quest and \projectq maintain a pure state in $2^n$ complex floating point numbers for a system of $n$ qubits, with (by default) double precision in each real and imaginary component; \quest can otherwise be configured to use single or quad precision. 
Both simulators store the state in \clang/\cpplang primitives, and so (by default) consume $16\times 2^n$\,B~\cite{MSDN_C_prim_sizes} in the state vector alone.
However \projectq incurs a $\times 1.5$ memory overhead during state allocation, and \quest clones the state vector in distributed applications.
Typical memory costs of both simulators on a single thread are shown in \figref~\ref{fig:quest_projq_memory}, which vary insignificantly from their multithreaded costs.
While \quest allows direct read and write access to the state-vector, \projectq's single amplitude fetching has a {Python overhead, and writing is only supported in batch which is memory expensive due to Python objects consuming more memory than a comparable C primitive - as much as $3\times$}~\cite{pythonDataObjects,python3xPrims}. 
{Iterating the state-vector in ProjectQ is therefore either very slow, or comes with an appreciable memory cost.}

\quest applies a single-qubit gate (a $2\times2$ matrix $G$) on qubit $q$ of an $N$-qubit \newedit{pure state-vector} $\ket{\psi} = \sum \limits_{n=0}^{2^N-1} \alpha_n \ket{n}$, represented as the complex vector $\vec{\alpha}$, by updating vector elements
\begin{align}
\begin{pmatrix}
\alpha_{n_i} \\
\alpha_{n_i + 2^q}
\end{pmatrix}
\mapsto
G
\begin{pmatrix}
\alpha_{n_i} \\
\alpha_{n_i + 2^q}
\end{pmatrix}
\end{align}
where $n_i = \left\lfloor i / {2^q} \right\rfloor 2^{q+1} + (i  \mod 2^q)$ for every integer $i \in [0, 2^{N-1} - 1]$.
This applies $G$ via $2^{N}$ computations of $a b + c d$ for complex $a, b, c, d$ and avoids having to compute and matrix-multiply a full $2^N \times 2^N$ unitary on the state-vector. This has a straight-forward generalisation to multi-control single-target gates, and lends itself to parallelisation.

\newedit{
We leverage the same hardware-optimised code to enact gates on $N$-qubit density matrices, by storing them as $2N$-qubit state-vectors,
}
\begin{align}
    \rho = \sum \limits_{j=0}^{2^N-1} \sum\limits_{k=0}^{2^N-1} \alpha_{j,k} \ket{j}\bra{k} 
        \ \ \rightarrow \ \ \rho^\prime=\sum \limits_{n=0}^{2^{2N}-1} \alpha_n' \ket{n}.
\end{align}
\newedit{
Here the object $\rho^\prime$ does not, in general, respect the constraint $\sum |\alpha^\prime_n|^2=1$. An operation $G_q \rho G_q^\dagger$, that is a gate on qubit $q$, can then be effected on $\rho^\prime$ as
$
    G_{q + N}^* G_q \rho^\prime,
$
by exploiting the Choi--Jamiolkowski isomorphism}
~\cite{choi1975completely} %
\newedit{This holds also for multi-qubit gates.
The distribution of the density matrix in this form lends itself well to the parallel simulation of dephasing and depolarising noise channels.
}

\subsubsection{Distributed}

How simulators partition the state vector between processes and communicate over the network is key to their performance on distributed memory architectures. All simulators we have found so far employ a simple partitioning scheme; the memory to represent a state vector is split equally between all processes holding that vector. A common strategy to then evaluate a circuit is to pair nodes such that upon applying a single qubit gate, every process must send and receive the entirety of its portion of the state vector to its paired process~\cite{parallel_qc_simmer_noisy,qhipster,distributed_quantum_plus_plus}.

\begin{figure}[t]
\centering
\hspace{.5cm} \begin{tikzpicture}[scale=.4]

\definecolor{ao}{rgb}{0.0, 0.5, 0.0}

\def\goodmemclonecol{ao!60!white}
\def\badmemclonecol{blue!60!white}

\def\goodmemcol{green!40!white}
\def\badmemcol{blue!40!white}
\def\overheadcol{red!40!white}
\def\emptycol{white}

\def\goodmembackcol{green!20!white}
\def\badmembackcol{blue!20!white}
\def\overheadbackcol{red!20!white}
\def\emptybackcol{white}

\def\goodmem{\draw[fill=\goodmemcol]}
\def\badmem{\draw[fill=\badmemcol]}
\def\overhead{\draw[fill=\overheadcol]}
\def\empty{\draw[fill=\emptycol]}

\def\goodmemback{\draw[fill=\goodmembackcol]}
\def\badmemback{\draw[fill=\badmembackcol]}
\def\overheadback{\draw[fill=\overheadbackcol]}
\def\emptyback{\draw[fill=\emptybackcol]}

\def\goodmemclone{\draw[pattern color=\goodmemclonecol, pattern=north east lines]}
\def\badmemclone{\draw[pattern color=\badmemclonecol, pattern=north east lines]}

\def\width{2}
\def\height{4}
\def\overheight{.3}
\def\dispr{.5}
\def\dispu{.6}
\def\dispgap{1.5}
\def\leggap{.25}
\def\legd{1}
\def\legsep{.25}
\def\textvgap{.75}
\def\texthgap{2*\width/4}
\def\legtexthgap{\legd/4}
\def\legtextvgap{\legd/2}

\def\frontthick{.6pt};
\def\backthick{.4pt};


\def\x{\dispr}; 
\def\y{\dispu};
\empty (\x,\y) rectangle (\x+\width,\y+\height);
\draw[line width=\backthick] (\x,\y) rectangle (\x+\width,\y+\height);

\def\x{0}; 
\def\y{0};
\badmem (\x,\y) rectangle  (\x+\width,\y+\height/2);
\overhead  (\x,\y+\height/2) rectangle (\x+\width,\y+\height/2+\overheight);
\empty  (\x,\y+\height/2+\overheight) rectangle (\x+\width,\y+\height);
\draw[line width=\frontthick] (\x,\y) rectangle (\x+\width,\y+\height);


\def\x{\width + \dispgap + \dispr}; 
\def\y{\dispu};
\badmemback (\x,\y) rectangle  (\x+\width,\y+\height/4);
\badmemback (\x,\y+\height/4) rectangle  (\x+\width,\y+\height/2);
\overheadback  (\x,\y+\height/2) rectangle (\x+\width,\y+\height/2+\overheight);
\emptyback  (\x,\y+\height/2+\overheight) rectangle (\x+\width,\y+\height);
\draw[line width=\backthick] (\x,\y) rectangle (\x+\width,\y+\height);

\def\x{\width + \dispgap}; 
\def\y{0};
\badmem (\x,\y) rectangle  (\x+\width,\y+\height/4);
\badmem (\x,\y+\height/4) rectangle  (\x+\width,\y+\height/2);
\badmemclone (\x,\y+\height/4) rectangle  (\x+\width,\y+\height/2);
\overhead  (\x,\y+\height/2) rectangle (\x+\width,\y+\height/2+\overheight);
\empty  (\x,\y+\height/2+\overheight) rectangle (\x+\width,\y+\height);
\draw[line width=\frontthick] (\x,\y) rectangle (\x+\width,\y+\height);

\node[draw=none] at (\x+\texthgap,\y+\height+\dispu+\textvgap) {$\times2$};


\def\x{2*\width + 2*\dispgap + \dispr}; 
\def\y{\dispu};
\goodmemback (\x,\y) rectangle  (\x+\width,\y+\height/2);
\goodmemback (\x,\y+\height/2) rectangle  (\x+\width,\y+3*\height/4);
\overheadback  (\x,\y+3*\height/4) rectangle (\x+\width,\y+3*\height/4+\overheight);
\emptyback  (\x,\y+3*\height/4+\overheight) rectangle (\x+\width,\y+\height);
\draw[line width=\backthick] (\x,\y) rectangle (\x+\width,\y+\height);

\def\x{2*\width + 2*\dispgap}; 
\def\y{0};
\goodmem (\x,\y) rectangle  (\x+\width,\y+\height/2);
\goodmem (\x,\y+\height/2) rectangle  (\x+\width,\y+3*\height/4);
\goodmemclone (\x,\y+\height/2) rectangle  (\x+\width,\y+3*\height/4);
\overhead  (\x,\y+3*\height/4) rectangle (\x+\width,\y+3*\height/4+\overheight);
\empty  (\x,\y+3*\height/4+\overheight) rectangle (\x+\width,\y+\height);
\draw[line width=\frontthick] (\x,\y) rectangle (\x+\width,\y+\height);

\node[draw=none] at (\x+\texthgap,\y+\height+\dispu+\textvgap) {$\times1.5$};


\def\x{3*\width + 3*\dispgap + \dispr}; 
\def\y{\dispu};
\goodmemback (\x,\y) rectangle  (\x+\width,\y+\height/2);
\goodmemback (\x,\y+\height/2) rectangle  (\x+\width,\y+\height/2+\height/8);
\overheadback  (\x,\y+\height/2+\height/8) rectangle (\x+\width,\y+\height/2+\height/8+\overheight);
\emptyback  (\x,\y+\height/2+\height/8+\overheight) rectangle (\x+\width,\y+\height);
\draw[line width=\backthick] (\x,\y) rectangle (\x+\width,\y+\height);

\def\x{3*\width + 3*\dispgap}; 
\def\y{0};
\goodmem (\x,\y) rectangle  (\x+\width,\y+\height/2);
\goodmem (\x,\y+\height/2) rectangle  (\x+\width,\y+\height/2+\height/8);
\goodmemclone (\x,\y+\height/2) rectangle  (\x+\width,\y+\height/2+\height/8);
\overhead  (\x,\y+\height/2+\height/8) rectangle (\x+\width,\y+\height/2+\height/8+\overheight);
\empty  (\x,\y+\height/2+\height/8+\overheight) rectangle (\x+\width,\y+\height);
\draw[line width=\frontthick] (\x,\y) rectangle (\x+\width,\y+\height);

\node[draw=none] at (\x+\texthgap,\y+\height+\dispu+\textvgap) {$\times 1.25$};


\def\x{4*\width + 4*\dispgap + \leggap}; 

\def\y{0}; 
\goodmem (\x,\y) rectangle (\x+\legd,\y+\legd);
\node[draw=none,anchor=west] at (\x+\legd+\legtexthgap,\y+\legtextvgap) {32 qubits};

\def\y{\legsep + \legd};
\badmem (\x,\y) rectangle (\x+\legd,\y+\legd);
\node[draw=none,anchor=west] at (\x+\legd+ \legtexthgap,\y+\legtextvgap) {31 qubits};

\def\y{2* \legsep + 2*\legd};
\badmem (\x,\y) -- (\x + \legd, \y) -- (\x + \legd, \y+\legd) -- cycle;
\badmemclone (\x,\y) -- (\x + \legd, \y) -- (\x + \legd, \y+\legd) -- cycle;
\goodmem (\x,\y) -- (\x, \y+ \legd) -- (\x + \legd, \y+\legd) -- cycle;
\goodmemclone (\x,\y) -- (\x, \y+ \legd) -- (\x + \legd, \y+\legd) -- cycle;
\node[draw=none,anchor=west] at (\x+\legd+ \legtexthgap,\y+\legtextvgap) {cloned};

\def\y{3*\legsep + 3*\legd};
\overhead(\x,\y) rectangle (\x+\legd,\y+\legd);
\node[draw=none,anchor=west] at (\x+\legd+ \legtexthgap,\y+\legtextvgap) {overhead};

\def\y{4*\legsep + 4*\legd};
\empty (\x,4* \legsep + 4*\legd) rectangle(\x + \legd, 4*\legsep +5* \legd);
\node[draw=none,anchor=west] at (\x+\legd+ \legtexthgap,\y+\legtextvgap) {free};

\end{tikzpicture} \\
\vspace{.4cm}
\includegraphics[width=.48\textwidth]{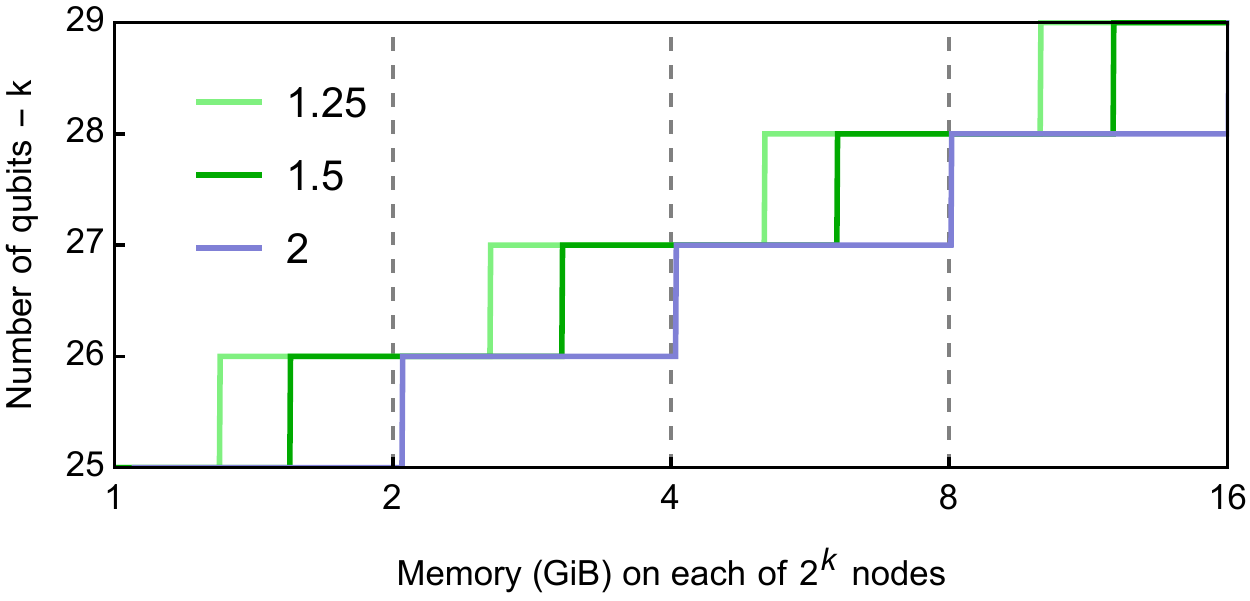}
\caption{An illustration of strategies to distribute the state vector between two 64\,GiB nodes. A complete cloning ($\times 2$ memory) of the partition on each node is wasteful. Half the partition can be cloned, at the cost of twice as many MPI messages, to fit another qubit into memory~\cite{parallel_qc_simmer_noisy}. Further division requires more communication for less memory overhead~\cite{qhipster}. The bottom plot shows the maximum number of qubits which can fit on $2^k$ nodes of varying memory, assuming a 50\,MiB overhead per node.
}
\label{fig:distributed_memory_schemes}
\end{figure}
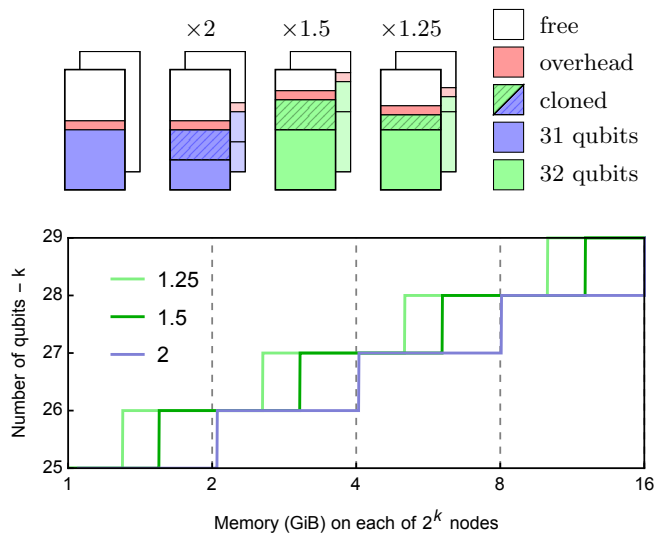

The number of communications between paired processes, the amount of data sent in each and the additional memory incurred on the compute nodes form a tradeoff. A small number of long messages will ensure that the communications are bandwidth limited, which leads to best performance in the communications layer. However this results in a significant memory overhead, due to the process having to store buffers for both the data it is sending and receiving, and in an application area so memory hungry as quantum circuit simulation this may limit the size of circuit that can be studied. On the other hand many short messages will minimise the memory overhead as the message buffers are small, but will lead to message latency limited performance as the bandwidth of the network fabric will not be saturated. This in turn leads to poor parallel scaling, and hence again limits the size of the circuit under consideration, but now due to time limitations. 
Note that the memory overhead is at most a factor 2, which due to the exponential scaling of the memory requirements, means only 1 less qubit may be studied. Some communication strategies and their memory overheads and visualised in \figref~\ref{fig:distributed_memory_schemes}.

\quest partitions the state vector equally between the processes within the job, and the message passing between the process pairs is so organised as to absolutely minimise the number of communications during the operation of a single gate. Thus parallel performance should be good, but there will be a significant memory overhead; in practice a factor of 2 as described above. For $n$ qubits distributed over $2^k$ nodes, these communications occur when operating on qubits with index $\ge n - k$, indexing from $0$.

An alternative strategy is to clone, send and receive only half of each node's data in two exchanges~\cite{parallel_qc_simmer_noisy}, incurring instead a $1.5\times$ memory cost. This often leaves room to simulate an additional qubit, made clear in \figref~\ref{fig:distributed_memory_schemes}. 
This strategy can be recursed further to reduce the memory overhead even more, and negligible additional memory cost can be achieved by communicating every amplitude separately as in~\cite{distributed_quantum_plus_plus}, though this comes at a significant communication cost, since a message passing pattern is latency dominated and will exhibit poor scaling with process count. 
However an improvement made possible by having two exchanges is to overlap the communication of the first message with the computation on the second half of the state vector, an optimisation implemented in \qhipster~\cite{qhipster}. This depends on the network effectively supporting asynchronous communications.


We also mention recent strategies for further reducing network traffic by optimising the simulated circuit through gate fusion, state reordering~\cite{qhipster,haener_45qb_sim} and rescheduling operations~\cite{haener_45qb_sim}, though opportunities for such optimisations may be limited.

In terms of the functionality implemented in the simulation packages we note that while \qhipster is limited to single and two-qubit controlled gates, \quest additionally allows the distributed operation of any-qubit controlled gates.


\subsubsection{\gpu}

Though \mpi distribution can be used for scalable parallelisation, networks are expensive and are overkill for deep circuits of few qubits. Simulations limited to 29 qubits can fit into a 12\,GB \gpu which offers high parallelisation at low cost. In our testing, \quest running a single Tesla K40m \gpu (retailing currently for $\sim$3.6\,k\,USD) outperforms $8$ distributed 12-core Xeon E5-2697 v2 series processors, currently retailing at $\sim$21\,k\,USD total, ignoring the cost of the network.

\quest is the first available simulator of both state-vectors and density matrices which can run on a \cuda enabled \gpu, {offering speedups of $\sim$5$\times$} over {already highly-parallelised 24-threaded single-node simulation}. We mention QCGPU~\cite{QCGPU} which is a recent \gpu-accelerated {single-node} simulator being developed with \newedit{Python} and OpenCL, and Quantumsim, a CUDA-based simulator of density matrices~\cite{quantumsimrepo}.

\subsubsection{Multi-platform}

{
QuEST is the only simulator which supports all of the above classical architectures. A simulation written in QuEST can be immediately deployed to all environments, from a laptop to a national-grade supercomputer, performing well at all simulation scales.
}
{We list the facilities supported by other state-of-the-art simulators in} \tabref \ref{tab:facility_table}.

\begin{table*}[htbp!]

{ 
\begin{tabular}{|c|c|c|c|c|c|}
\hline
Simulator & multithreaded & distributed & GPU accelerated & stand-alone & density matrices \\
\hline
\quest & \checkmark & \checkmark & \checkmark & \checkmark & \checkmark \\
\hline 
\qhipster & \checkmark & \checkmark & & \checkmark & \\
\hline
\quantumpp & \checkmark &  & & ? & \\
\hline
Quantumsim & & & \checkmark & & \checkmark \\
\hline
QCGPU & \checkmark & & \checkmark & & \\
\hline 
\projectq & \checkmark & & & &\\
\hline
\end{tabular}
\caption{{A comparison of the facilities offered by some publicly available, state-of-the-art simulators. Note the distributed adaptation of Quantum++}\cite{distributed_quantum_plus_plus} {is not currently publicly available}.
Here, density matrices refers to the ability to precisely represent mixed states.
}
} 

\label{tab:facility_table}
\end{table*}


\subsection{Algorithm}

\begin{figure}[t]
\centering
\input{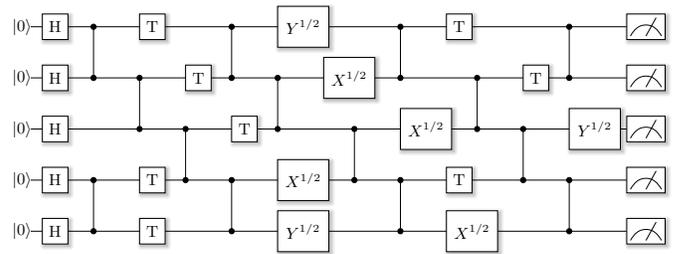}
\caption{An example of a depth 10 random circuit on 5 qubits, of the linear topology described in \cite{rand_quant_circuits}. This diagram was generated using \projectq's circuit drawer.}
\label{fig:randcircexample}
\end{figure}

We compare \quest and \projectq performing simulations of universal psuedo-random quantum circuits of varying depth and number of qubits. 
A random circuit contains a random sequence of gates, in our case with gates from the universal set $\{H$, $T$, $C(Z)$, $X^{1/2}$, $Y^{1/2}\}$. These are the Hadamard, $\pi/8$, controlled-phase and root Pauli X and Y gates. Being computationally hard to simulate, random circuits are a natural algorithm for benchmarking simulators~\cite{rand_quant_circuits}. 
We generate our random circuits by the algorithm in~\cite{rand_quant_circuits}, which fixes the topology for a given depth and number of qubits, though randomises the sequence of single qubit gates. An example is shown in \figref~\ref{fig:randcircexample}. The total number of gates (single plus control) goes like $\mathcal{O}(n d)$ for an $n$ qubit, depth $d$ random circuit, and the ratio of single to control gates is mostly fixed at $1.2 \pm 0.2$, so we treat these gates as equal in our runtime averaging. For measure, a depth 100 circuit of 30 qubits features 1020 single qubit gates and 967 controlled phase gates.

Though we here treat performance in simulating a random circuit as an indication of the general performance of the simulator, we acknowledge that specialised simulators may achieve better performance on particular classes of circuits. For example, \projectq can utilise topological optimisation or classical emulation to shortcut the operation of particular subcircuits, such as the quantum Fourier transform~\cite{projq_emulator_abilities}.

We additionally study \quest's communication efficiency by measuring the time to perform single qubit rotations on distributed hardware.


\section{Setup}


\subsection{\label{sec:hardware}Hardware}

We evaluate the performance of \quest and \projectq using Oxford's computing facilities, specifically the \arcus supercomputer, and the UK National Supercomputing facility \archer.


\quest and \projectq are compared on single nodes with 1-16 threads, on \arcus
with nodes of 64, 128 and 256 GiB memory (simulating 1-31, 32 and 33 qubits respectively), each with two 8-core Intel Xeon E5-2640 V3 processors and a collective last level cache (\llc) size of 41\,MB between two \numa banks.
We furthermore benchmark \quest on \arcus Tesla K40m \gpu nodes, which with 12\,GB global memory over 2880 \cuda cores, can simulate up to 29 qubit circuits.


\quest and \projectq are also compared on \archer,
a CRAY XC30 supercomputer. \archer contains both 64 and 128 GiB compute nodes, each with two 12-core Intel Xeon E5-2697 v2 series processors linked by two QuickPath Interconnects, and a collective \llc of 61\,MB between two \numa banks. Thus a single node is capable of simulating up to 32 qubits with 24 threads.
We furthermore evaluate the scalability of \quest when distributed over up to 2048 \archer compute nodes, linked by a Cray Aries interconnect, which supports an MPI latency of $\sim 1.4\pm 0.1\,\mu$s and a bisection bandwidth of 19\,TB/s.


\subsection{Software}

\subsubsection{Installation}


On \arcus, we compile both single-node \quest v0.10.0 and \projectq's \cpplang backend with \gcc 5.3.0, which supports \openmp 4.0~\cite{openmp_versions} for parallelisation among threads.
For \gpu use, \questGPU v0.6.0 is compiled with the NVIDIA \cuda 8.0.
\projectq v0.3.5 is run with \python 3.5.4, inside an \anaconda 4.3.8 environment.


On \archer, \projectq v0.3.6 is compiled with \gcc 5.3.0, and run in \python 3.5.3 inside an \anaconda 4.0.6.
\quest is compiled with \icc 17.0.0 which supports \openmp 4.5 \cite{openmp_versions}, and is distributed with the MPICH3 implementation of the \mpi 3.0 standard, optimised for the Aries interconnect.



\subsubsection{\label{sec:projq_config}Configuration}

We attempt to optimise \projectq when simulating many qubits by enabling gate fusion only for multithreaded simulations~\cite{projq_cache_anomaly_haener}.

\begin{codebox}
\begin{minted}{python}
from projectq import MainEngine
from projectq.backends import Simulator

MainEngine(
    backend=Simulator(
        gate_fusion=(threads > 1)))
\end{minted}
\end{codebox}

We found that \projectq's multithreaded simulation of few qubit random circuits can be improved by disabling all compiler engines, to reduce futile time spent optimising the circuit in \python.

\begin{codebox}
\begin{minted}{python}
MainEngine(
    backend=Simulator(gate_fusion=True),
    engine_list=[])
\end{minted}
\end{codebox}

However, this disables \projectq's ability to perform classical emulation and gate decomposition, and so is not explored in our benchmarking.
We studied \projectq's performance for different combinations of compiler engines, number of gates considered in local optimisation and having gate fusion enabled, and found the above configurations gave the best performance for random circuits on our tested hardware.

Our benchmarking measures the runtime of strictly the code responsible for simulating the sequence of gates, and excludes the time spent allocating the state vector, instantiating or freeing objects or other one-time overheads. 

In \projectq, this looks like:

\begin{codebox}
\begin{minted}{python}
# prepare the simulator
sim = Simulator(gate_fusion=(threads > 1))
engine = MainEngine(backend=sim)
qubits = engine.allocate_qureg(num_qubits)
engine.flush()

# ensure we're in the 0 state
sim.set_wavefunction([1], qubits)
sim.collapse_wavefunction(
    qubits, [0]*num_qubits)
engine.flush()

# start timing, perform circuit

# ensure cache is empty
engine.flush()
sim._simulator.run()

# stop timing
\end{minted}
\end{codebox}

and in \quest:

\begin{codebox}
\begin{minted}{cpp}
// prepare the simulator
QuESTEnv env = createQuESTEnv();
Qureg qubits = createQureg(num_qubits, env);

// ensure we're in the 0 state
initZeroState(&qubits);

// start timing, perform circuit

// ensure distributed work finishes
syncQuESTEnv(env);

// stop timing
\end{minted}
\end{codebox}




\section{Results}



\subsection{Single Node Performance}

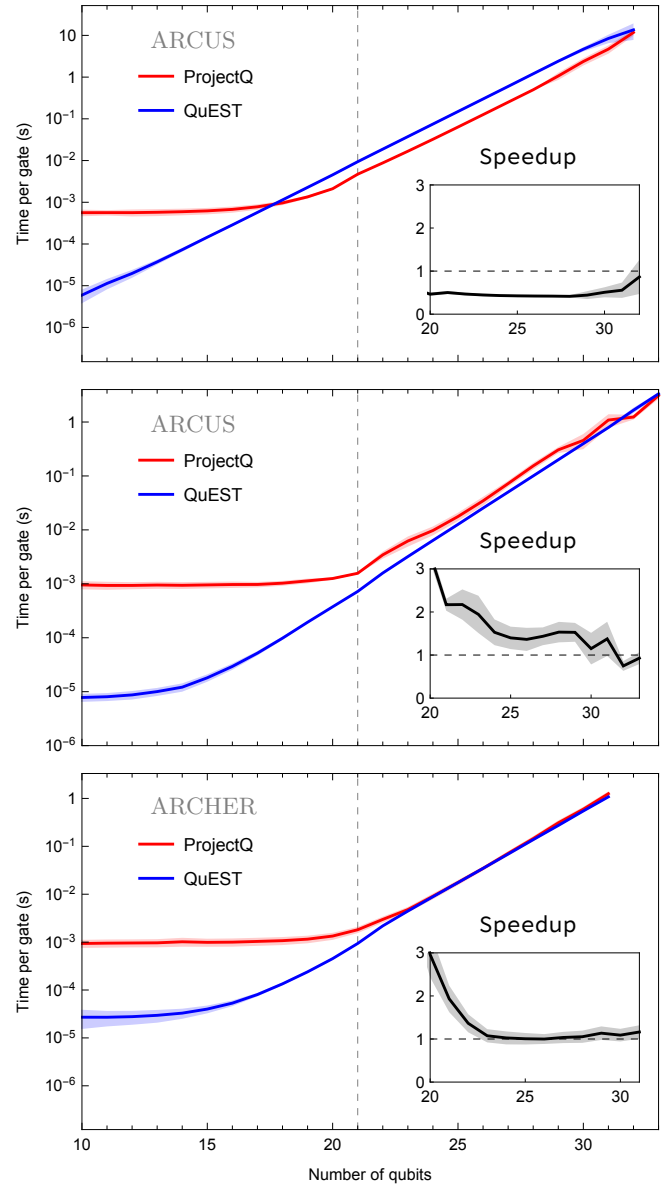
\begin{figure}[b]
\centering

\begin{subfigure}
\centering
\begin{tikzpicture}
\draw (0, 0) node[inner sep=0] {\includegraphics[width=.48\textwidth,trim={0 20 0 0},clip]{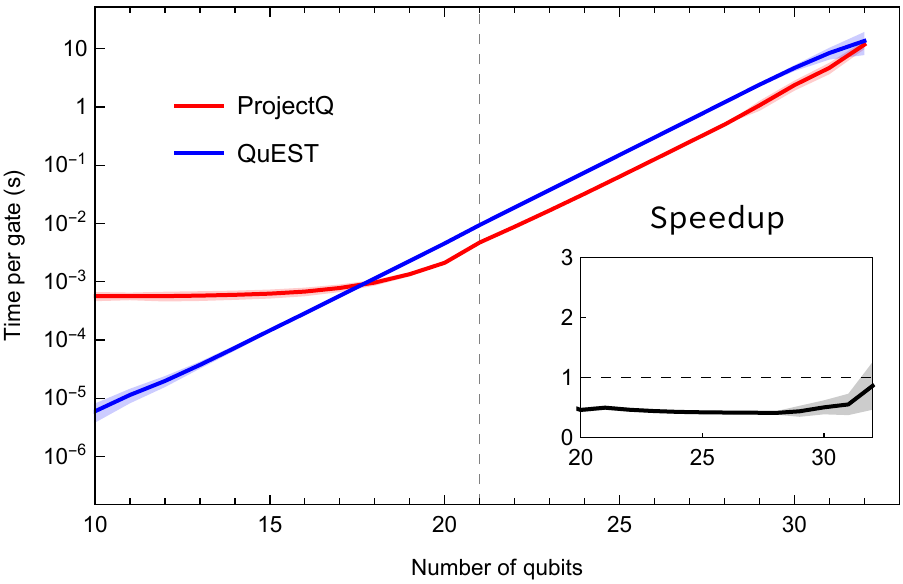}};
\node[draw=none, text=gray, anchor=west] at (-2.6, 1.92) {\small ARCUS};
\end{tikzpicture}
\end{subfigure}
\begin{subfigure}
\centering
\begin{tikzpicture}
\draw (0, 0) node[inner sep=0] {\includegraphics[width=.48\textwidth,trim={0 20 0 0},clip]{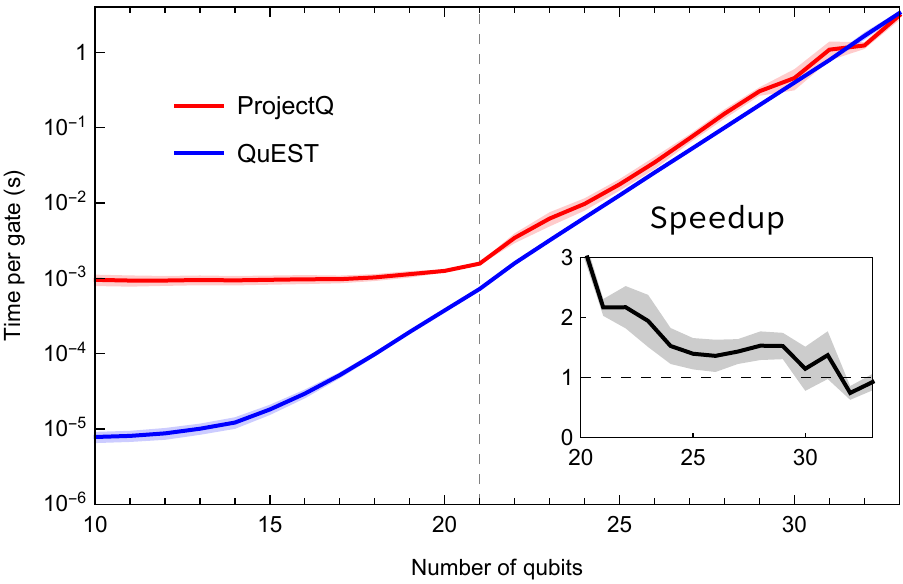}};
\node[draw=none, text=gray, anchor=west] at (-2.6, 1.92) {\small ARCUS};
\end{tikzpicture}
\end{subfigure}
\begin{subfigure}
\centering
\begin{tikzpicture}
\draw (0, 0) node[inner sep=0] {\includegraphics[width=.48\textwidth]{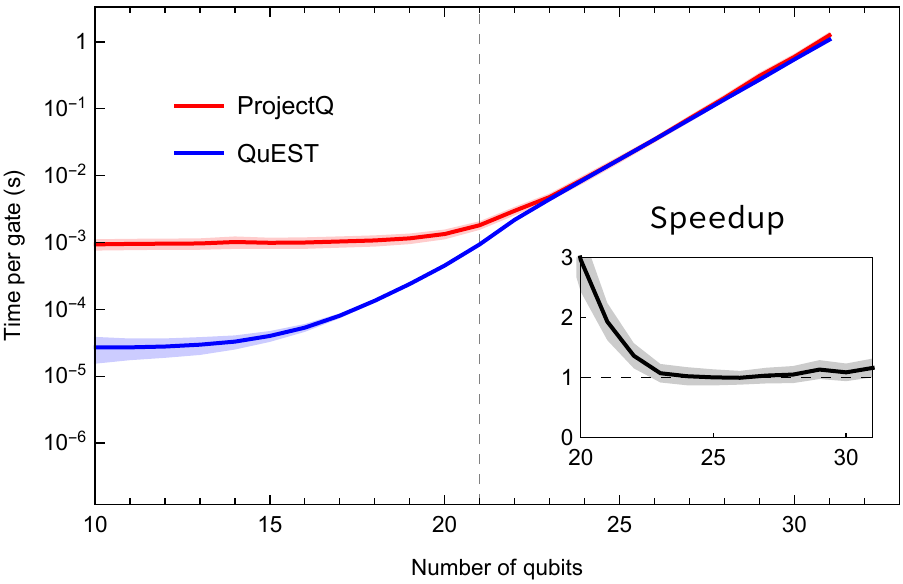}};
\node[draw=none, text=gray, anchor=west] at (-2.6, 2.28) {\small \archer};
\end{tikzpicture}
\end{subfigure}%

\caption{Comparison of \quest and \projectq when simulating random circuits over 1, 16 (on \arcus) and 24 (on \archer) threads (top to bottom). Coloured lines indicate the mean, with shaded regions indicating a standard deviation either side, over a total of $\sim$77\,k simulations of varying depth. Vertical dashed lines indicate the maximum number of qubits for which the entire state vector fits into the \llc. The speedup subgraphs show the ratio of \projectq to \quest runtime.}
\label{fig:quest_projq_runtime_comp}
\end{figure}

\begin{figure}[t]
\centering
\includegraphics[width=.48\textwidth]
{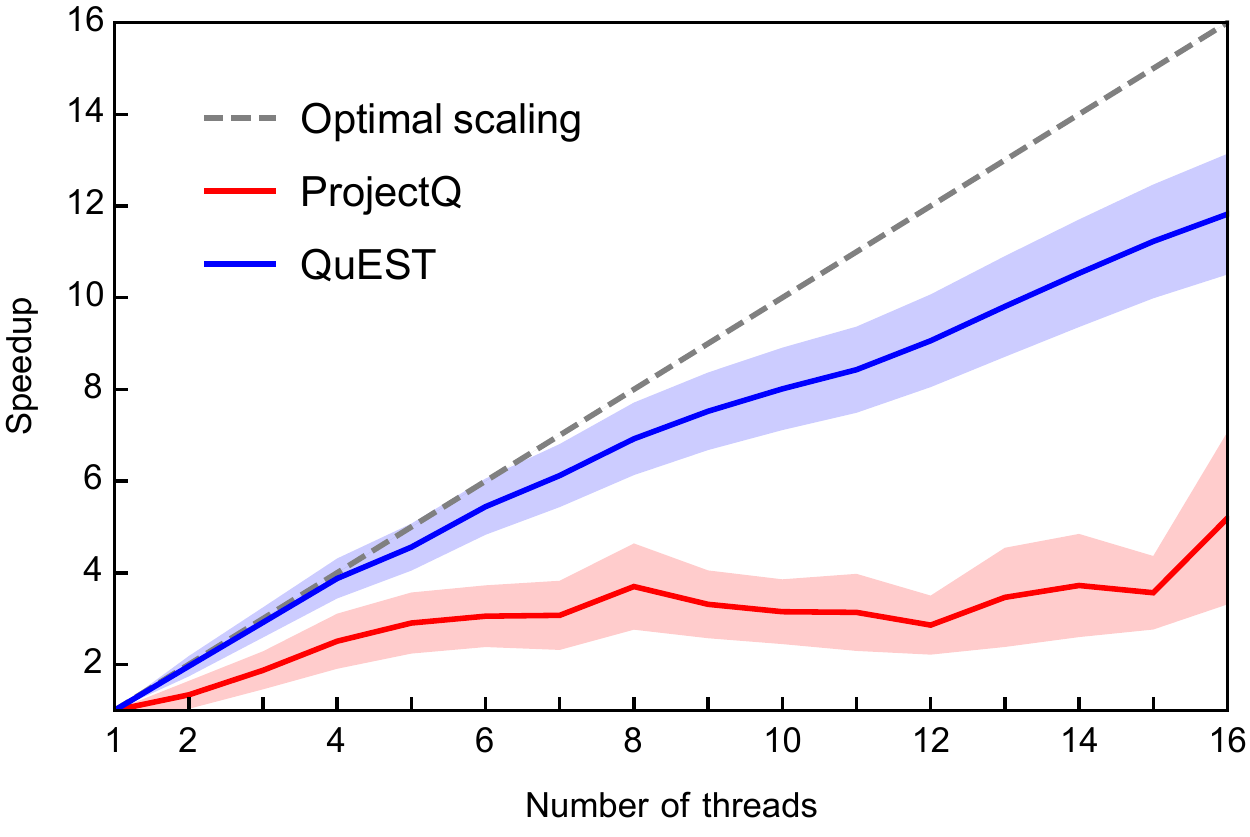}
\caption{
Single-node strong scaling achieved when parallelising (through \openmp) 30 qubit random circuits across a varying number of threads on a 16-\cpu{} \arcus compute node. Solid lines and shaded regions indicate the mean and a standard deviation either side (respectively) of $\sim$7\,k simulations of circuit depths between 10 and 100.
}
\label{fig:quest_projq_thread_scaling}
\end{figure}

\begin{figure}[t]
\centering
\includegraphics[width=.48\textwidth]{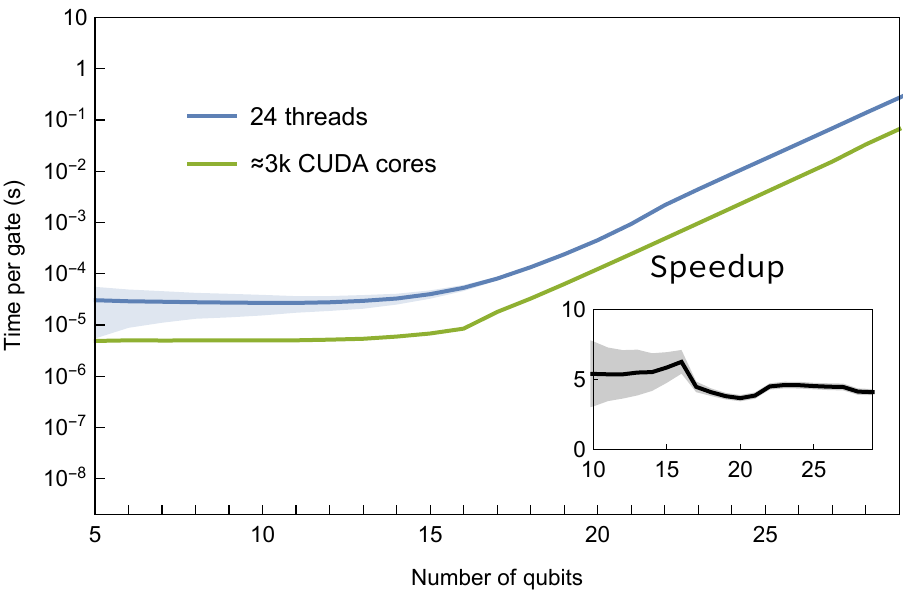}
\caption{\quest's single-node performance using multithreading and \gpu acceleration to parallelise random circuit simulations. The subplot shows the speedup (ratio of runtimes) that a \gpu of 2880 \cuda cores on \arcus achieves against 24 threads on \archer.}
\label{fig:quest_gpu}
\end{figure}

The runtime performance of \quest and \projectq, presented in \figref~\ref{fig:quest_projq_runtime_comp}, varies with the architecture on which they are run, and the system size they simulate. Anomalous slowdown of \projectq at 22 qubits may be explained by the \llc becoming full, due to its use of cache blocking through gate fusion~\cite{projq_cache_anomaly_haener}.

For fewer than $\sim$22 qubits, \projectq's \python overhead is several orders of magnitude slower than \quest's \clang overhead, independent of circuit depth. The \python overhead can be reduced by disabling some simulation facilities - see Section~\ref{sec:projq_config}.
For larger systems, the time spent in \projectq's \cpplang backend operating on the state vector dominates total runtime, and the time per gate of both simulators grows exponentially with increasing number of qubits. 

On a single ARCUS-B thread, \projectq becomes twice as fast as \quest, attributable to its sophisticated circuit evaluation. However, these optimisations appear to scale poorly; \quest outperforms \projectq on 16 threads on ARCUS-B, and on ARCHER both simulation packages are equally fast on 24 threads.
This is made explicit in the strong scaling over threads shown in
\figref~\ref{fig:quest_projq_thread_scaling}, which reveals \projectq's scaling is not monotonic. Performance suffers with the introduction of more than 8 threads, though is restored at 16.



We demonstrate \quest's utilisation of a \gpu for highly parallelised simulation in \figref~\ref{fig:quest_gpu}, achieving a speedup of $\sim$5$\times$ from \quest and \projectq on 24 threads.


\subsection{Distributed Performance}

\begin{figure}[b]
\centering
\includegraphics[width=.48\textwidth]{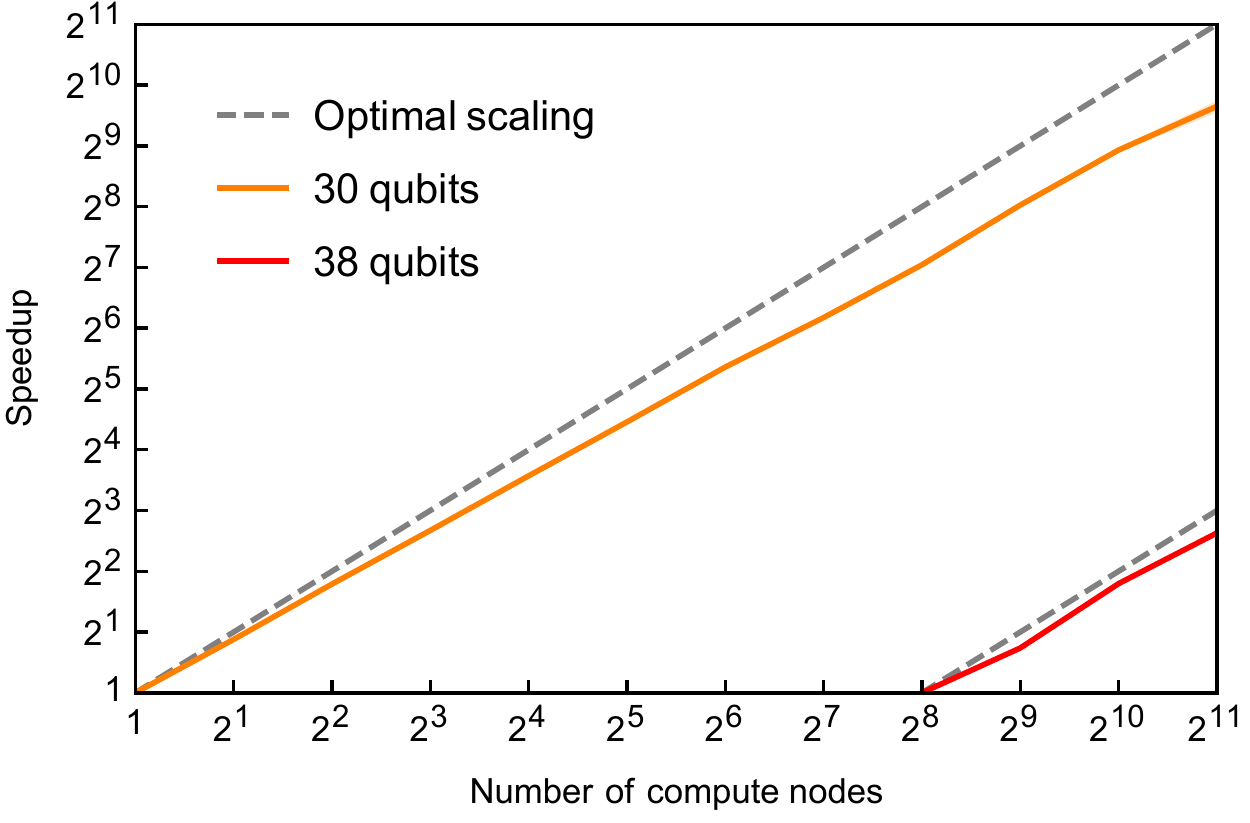}
\caption{\quest multinode strong scaling when distributing (through \mpi) a depth 100 (depth 10) random circuit simulation of 30 qubits (38 qubits) across many 24-thread 64\,GiB \archer nodes. 
}
\label{fig:quest_mpi_rc_strong_scaling}
\end{figure}

Strong scaling of \quest simulating a 30 and 38 qubit random circuit, distributed over 1 to 2048 \archer nodes, is shown in \figref~\ref{fig:quest_mpi_rc_strong_scaling}. In all cases one MPI process per node was employed, each with 24 threads. 
Recall that \quest's communication strategy involves cloning the state vector partition stored on each node.
The 30 qubit (38 qubit) simulations therefore demand 32\,GiB (8\,TiB) memory (excluding overhead), and require at least 1 node (256 nodes), whereas \qhipster{}'s strategy would fit a 31 qubit (39 qubit) simulation on the same hardware~\cite{qhipster}.

\begin{figure}[htbp!]
\centering
\includegraphics[width=.48\textwidth]{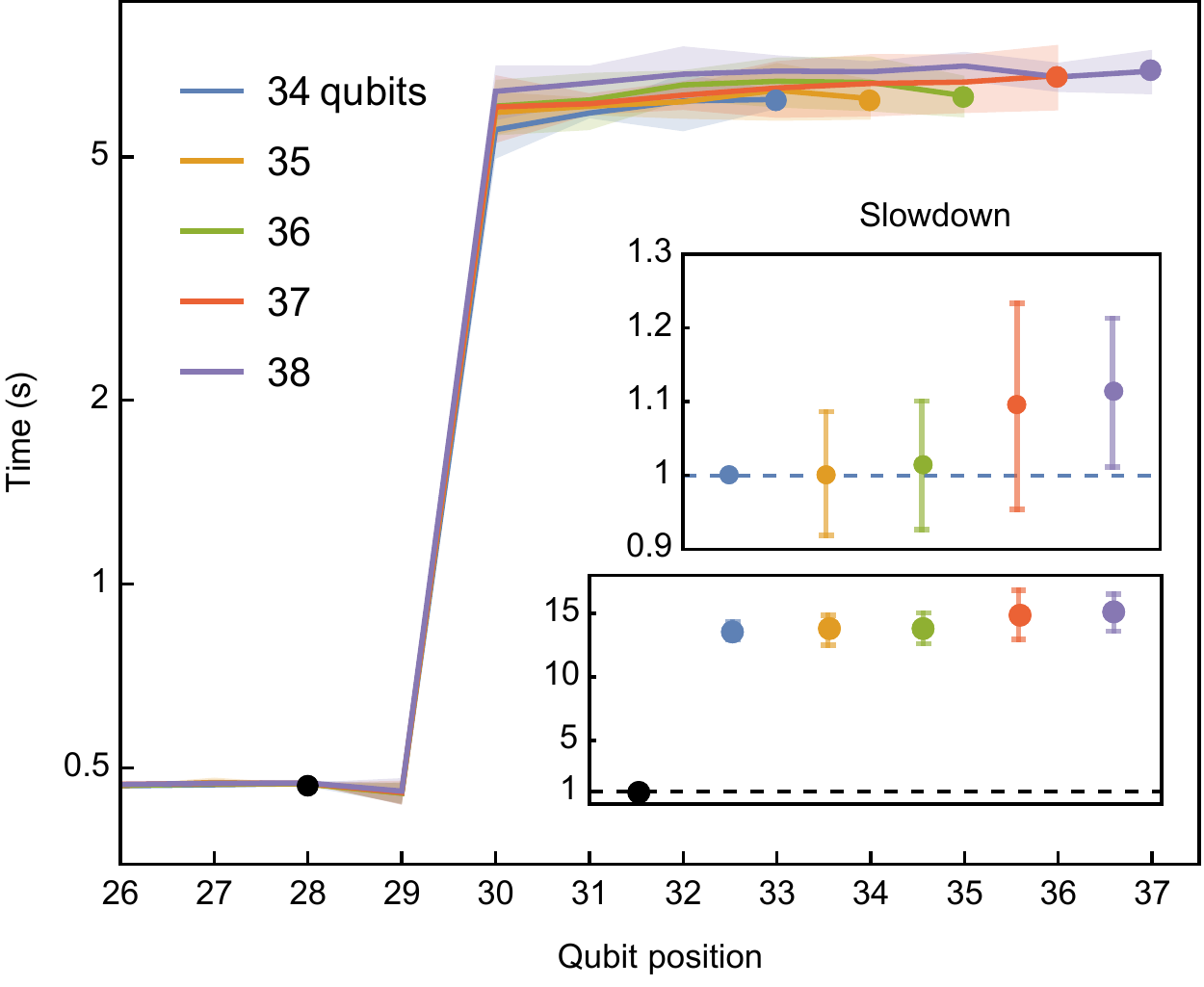}
\caption{
\quest multinode weak scaling of a single qubit rotation, distributed on \{16, 32, 64, 128, 256\} \archer nodes respectively, each with 24 threads between two sockets and 64\,GiB of memory. Communication occurs for qubits at positions $\ge 30$, indexing from 0. Time to rotate qubits at positions 0-25 are similar to those of 26-29 and are omitted. The bottom subplot shows the slowdown caused by communication, while the top subplot shows the slowdown of rotating the final (communicated) qubit as the total number of qubits simulated increases from 34. 
}
\label{fig:quest_weak_rotation_scaling}
\end{figure}

Communication cost is shown in \figref~\ref{fig:quest_weak_rotation_scaling} as the time to rotate a single qubit when using just enough nodes to store the state-vector; the size of the partition on each node is constant for increasing nodes and qubits. \quest shows excellent weak scaling, and moving from 34 to 37 simulated qubits slows \quest by a mere $\approx 9\%$. It is interesting to note that the  equivalent results for \qhipster show a slowdown of $\approx 148\%$ \cite{qhipster}, but this is almost certainly a reflection of the different network used in generating those results, rather than in any inherent weakness in \qhipster itself. 
\quest and \qhipster show comparable $\sim 10^1$ slowdown when operating on qubits which require communication against operations on qubits which do not (shown in the bottom subplot of \figref~\ref{fig:quest_weak_rotation_scaling}). Though such slowdown is also network dependent, it is significantly smaller than the $\sim 10^6$ slowdown reported by the \quantumpp adaptation on smaller systems \cite{distributed_quantum_plus_plus}, and reflects a more efficient communication strategy.
We will discuss these network and other hardware dependencies further in future work, and also intend to examine \qhipster on \archer so a true like with like comparison with \quest can be made.




\section{Summary}

This paper introduced \quest, a new high performance open source framework for simulating universal quantum computers. We demonstrated \quest shows good strong scaling over \openmp threads, competitive with a state of the art single-node simulator \projectq when performing multithreaded simulations of random circuits. We furthermore parallelised \quest on a \gpu for a $5 \times$ speedup over a 24 threaded simulation, and a $40 \times$ speedup over single threaded simulation.
\quest also supports distributed memory architectures via message passing with \mpi, and we've shown \quest to have excellent strong and weak scaling over multiple nodes. This behaviour has been demonstrated for up to 2048 nodes and has been used to simulate a 38 qubit random circuit.
Despite its relative simplicity, we found \quest's communication strategy yields comparable performance to \qhipster's, and strongly outperforms the distributed adaptation of \quantumpp. \quest can be downloaded in Reference~\cite{quest_site}

\section{Acknowledgements}

The authors would like to thank Mihai Duta as a contributor to \quest, and the \projectq team for helpful advice in configuring \projectq on \arcus and \archer.
The authors are grateful to the NVIDIA corporation for their donation of a Quadro P6000 to further the development of \quest.
The authors also acknowledge the use of the University of Oxford Advanced Research Computing (ARC) facility (\url{http://dx.doi.org/10.5281/zenodo.22558}) and the ARCHER UK National Supercomputing Service (\url{http://www.archer.ac.uk}) in carrying out this work. TJ thanks the Clarendon Fund for their support. SCB acknowledges EPSRC grant EP/M013243/1, and further acknowledges US funding with the following statement:
The research is based upon work supported by the Office
of the Director of National Intelligence (ODNI), Intelligence
Advanced Research Projects Activity (IARPA), via
the U.S. Army Research Office Grant No. W911NF-16-1-0070. 
The views and conclusions contained herein are those
of the authors and should not be interpreted as necessarily
representing the official policies or endorsements, either expressed
or implied, of the ODNI, IARPA, or the U.S. Government.
The U.S. Government is authorized to reproduce and
distribute reprints for Governmental purposes notwithstanding
any copyright annotation thereon. Any opinions, findings,
and conclusions or recommendations expressed in this material
are those of the author(s) and do not necessarily reflect the
view of the U.S. Army Research Office.


\bibliography{bibliography.bib}

\end{document}